\documentclass[aps,reprint,nofootinbib]{revtex4-2}

\usepackage{graphicx,amsmath,amssymb,latexsym,tikz}

\newcommand{\numberin}[1]{\#(#1)}

\newcommand{\ket}[1]{\left | #1 \right \rangle}
\newcommand{\bra}[1]{\left \langle #1 \right |}

\newcommand{\amp}[2]{\left \langle #1 | #2 \right \rangle}

\newcommand{\ptest}[1]{\boldsymbol{\Pi}_{#1}}
\newcommand{\ident}{\boldsymbol{1}}

\newcommand{\myrhomb}[2]{\draw[fill=white] (-1+#1,0+#2)--(0+#1,0.4+#2)--(1+#1,0+#2)--(0+#1,-0.4+#2)--cycle;}

\begin{document}

\title{Quantum paradoxical knowledge}

\author{Benjamin Schumacher}
\email[Corresponding author: ]{schumacherb@kenyon.edu}
\affiliation{Department of Physics, Kenyon College, Gambier, OH 43022}

\author{Michael D. Westmoreland}
\affiliation{Department of Mathematics, Denison University, Granville, OH 43023}

\begin{abstract}
	We generalize the quantum ``pigeonhole paradox'' to quantum paradoxes
	involving arbitrary types of particle relations, including orderings,
	functions and graphs.
\end{abstract}

\maketitle

\section{Introduction}

Aharonov et al. \cite{qpigeonhole} showed that quantum particles
can apparently violate the ``pigeonhole principle''.  Ordinarily, if three 
particles (``pigeons'') are placed in two boxes (``pigeonholes''), 
at least two of the particles must occupy the same box.  
However, in a quantum scenario involving both state preparation
and post-selection, we may be able to infer with certainty
that no particular pair is found occupying the same box.

An example illustrates the idea.  Each particle may 
occupy either the left or right boxes (states $\ket{L}$ and $\ket{R}$),
or of course any superposition of these.  For example, a particle
may be in states
\begin{equation}
\begin{split}
	\ket{\pm} & = \frac{1}{\sqrt{2}}  \left ( \ket{L} \pm \ket{R} \right), \\
	\ket{\pm i} & = \frac{1}{\sqrt{2}}  \left ( \ket{L} \pm i \ket{R} \right),
\end{split}
\end{equation}
and so on.  We prepare the three particles in the product state
\begin{equation}
	\ket{\Psi} = \ket{+_1, +_2, +_3} .
\end{equation}

At an intermediate time, one of three possible projective 
measurements is performed, testing whether a given pair 
of particles occupy the same box.  
The projections for these measurements may be
written $\ptest{12}$, $\ptest{23}$ and $\ptest{13}$.   
These projections commute, and in this sense
the observables are compatible.  However, the 
physical procedure for testing $\ptest{12}$ only
is distinct from the procedures for $\ptest{23}$ only
or $\ptest{13}$ only, and so they are complementary
measurements.
If we measure the pair $(i,j)$ and find an affirmative
answer, the (non-normalized) resulting state is 
$\ptest{ij} \ket{\Psi}$.

Finally, a measurement is made
on all three particles, and we post-select on the eigenstate
\begin{equation}
	\ket{\Phi} = \ket{+i_1, +i_2, +i_3}.
\end{equation}
This final outcome is possible no matter what intermediate measurement
was made.  However, for any pair $i$ and $j$ we find that
\begin{equation}
	\bra{\Phi} \ptest{ij} \ket{\Psi} = 0 .
\end{equation}
Therefore, in those cases where the final state is found to be 
$\ket{\Phi}$, we are in a position to know with certainty that
particles $i$ and $j$ were not found in the same box, for any
distinct values of $i$ and $j$.

The preparation of $\ket{\Psi}$ and post-selection of $\ket{\Phi}$
give us knowledge about pairs of particles that cannot be
reconciled with any particular distribution of the particles among
the boxes.  We may say that we have {\em paradoxical
knowledge} of the relations among the particles in the boxes.
Here we generalize the idea of paradoxical knowledge to
other types of relations including orderings, functions and  graphs,
and we show how such knowledge may arise in quantum 
systems.
Our results provide a new framework for examining the
non-classical information provided by quantum measurements
and suggest new experiments for small quantum computers.

\section{Framework}

We begin by outlining a simplified framework for our analysis.
There is an underlying finite set $P$ of objects called particles,
which we denote $1, 2, \ldots, N$.  A binary relation $r$ on $P$ 
is a subset of $P \times P$, and we say that $i$ and $j$ have
the $r$-relation provided $(i,j) \in r$.  We also specify a particular
collection $R$ of relevant binary relations on the particles.
For example, we may wish to restrict our attention to total
orderings on P, or to equivalence relations on P, or to some other class.

We have {\em knowledge} of the relation if, for some pairs
$(i,j)$, we can either affirm or deny with certainty that $i$ and $j$
are found to be related.  We can describe this by disjoint sets 
of pairs $A$ and $D$.  If $(i,j) \in A$, the affirmable set, 
then we are certain that $i$ is related to $j$; if $(i,j) \in D$, 
the deniable set, we are certain that $i$ is not related to $j$.

Our knowledge is {\em paradoxical} if it is not consistent with any
particular relation in $R$.  That is, $[A,D]$ represents paradoxical
knowledge if there does not exist $r \in R$ such that
\begin{itemize}
	\item  $(i,j) \in A$ implies that $(i,j) \in r$, and
	\item $(i,j) \in D$ implies that $(i,j) \notin r$ .
\end{itemize}
The sets $A$ and $D$ represent a partial description of some binary
relation on $P$, and this is paradoxical if the relation cannot 
be in $R$.

How can we arrive at quantum paradoxical knowledge?
Each relation $r \in R$ is associated with a quantum state
$\ket{r}$ of our system.
We prepare a fixed initial state $\ket{\Psi}$ that is a 
uniform superposition of all of the relevant relations:
\begin{equation}
	\ket{\Psi} = \sum_{r \in R} \ket{r} .
\end{equation}
(For convenience, we ignore normalization of our states,
since we do not need to calculate probabilities other than 0 or 1.)

At an intermediate time, we make a projective measurement
of whether a particular pair $(i,j)$ is related.  The projection
has the property that
\begin{equation}
	\ptest{ij} \ket{r} = \left \{ 
		\begin{array}{cl}
		\ket{r} & (i,j) \in r \\ 0 & (i,j) \notin r 
		\end{array}
		\right.
\end{equation}
As before, all such projections commute, but a measurement
of $\ptest{ij}$ only is complementary to measurements for
other possible pairs.

Finally, we make a measurement on the system, one of whose eigenstates
is $\ket{\Phi}$.  This can be written (in ``bra'' form) as
\begin{equation}
	\bra{\Phi} = \sum_{r \in R} \phi_{r} \bra{r} ,
\end{equation}
for some coefficients $\phi_{r}$.  
We post-select on the final result $\ket{\Phi}$ for this measurement.

Suppose our intermediate measurement is $\ptest{ij}$.  What
sort of knowledge does our preparation and post-selection
procedure provide?  We note that 
$\bra{\Phi} \ptest{ij} \ket{\Psi} + \bra{\Phi} (\ident - \ptest{ij}) \ket{\Psi} = \amp{\Phi}{\Psi}$.
We therefore require that $\amp{\Phi}{\Psi} \neq 0$,
so that the post-selected result $\ket{\Phi}$ is always possible given
the choice of $\ptest{ij}$.  There are three possibilities:
\begin{itemize}
	\item  If $\bra{\Phi} \ptest{ij} \ket{\Psi} = \amp{\Phi}{\Psi} \neq 0$, but
		$\bra{\Phi} (\ident - \ptest{ij}) \ket{\Psi} = 0$, then we
		can affirm with certainty that $i$ and $j$ were found
		to be related.  Then $(i,j) \in A$, the affirmable set.
	\item  If $\bra{\Phi} \ptest{ij} \ket{\Psi} = 0$, but
		$\bra{\Phi} (\ident - \ptest{ij}) \ket{\Psi} = \amp{\Phi}{\Psi} \neq 0$, then we
		can deny with certainty that $i$ and $j$ were found
		to be related.  Then $(i,j) \in D$, the deniable set.
	\item  If both $\bra{\Phi} \ptest{ij} \ket{\Psi} \neq 0$ and
		$\bra{\Phi} (\ident - \ptest{ij}) \ket{\Psi} \neq 0$, then
		we cannot infer the intermediate measurement result
		with certainty, and so $(i,j)$ is in neither $A$ nor $D$.
\end{itemize}

We illustrate our framework via the pigeonhole example.
In our simplified scheme, only one system state 
is associated with each relation $r$.
This is different from the original quantum pigeonhole construction,
in which different states---e.g., $\ket{L,L,R}$ and $\ket{R,R,L}$---have
identical togetherness relations among the particles.  Nevertheless,
we can capture the essential idea.

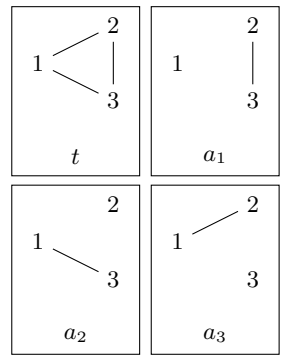
\begin{figure}
	\begin{center}
	\begin{tabular}{cc}
        	\begin{tikzpicture}[scale=1.0]
        		\node (A) at (0,0) {$1$};
        		\node (B) at (1,0.5) {$2$};
        		\node (C) at (1,-0.5) {$3$};
        		\node (R) at (0.5,-1.25) {$t$};
        		\draw (A) -- (B);
        		\draw (B) -- (C);
        		\draw (C) -- (A);
		\draw (-0.35,-1.5) rectangle (1.35,0.75);
	\end{tikzpicture} 
	& 
	\begin{tikzpicture}[scale=1.0]
        		\node (A) at (0,0) {$1$};
        		\node (B) at (1,0.5) {$2$};
        		\node (C) at (1,-0.5) {$3$};
        		\node (R) at (0.5,-1.25) {$a_1$};
        		\draw (B) -- (C);
		\draw (-0.35,-1.5) rectangle (1.35,0.75);
	\end{tikzpicture} 
	\\
	\begin{tikzpicture}[scale=1.0]
        		\node (A) at (0,0) {$1$};
        		\node (B) at (1,0.5) {$2$};
        		\node (C) at (1,-0.5) {$3$};
        		\node (R) at (0.5,-1.25) {$a_2$};
        		\draw (C) -- (A);
		\draw (-0.35,-1.5) rectangle (1.35,0.75);
	\end{tikzpicture} 
	& 
	\begin{tikzpicture}[scale=1.0]
        		\node (A) at (0,0) {$1$};
        		\node (B) at (1,0.5) {$2$};
        		\node (C) at (1,-0.5) {$3$};
        		\node (R) at (0.5,-1.25) {$a_3$};
        		\draw (A) -- (B);
		\draw (-0.35,-1.5) rectangle (1.35,0.75);
	\end{tikzpicture}
	\end{tabular}
	\end{center}
	\caption{Togetherness relations for three particles in 
		two boxes.  \label{fig-pigeonhole}}
\end{figure}

The four relevant togetherness relations for three particles
are shown in Figure~\ref{fig-pigeonhole}.
In relation $t$, all particles are in the same box.
In relation $a_{i}$, particle $i$ is alone in one box and 
the other two particles are together in the other box. 
Thus, for example, we find that 
$\ptest{12} \ket{t} = \ket{t}$ and $\ptest{12} \ket{a_{3}} = \ket{a_{3}}$, 
but $\ptest{12} \ket{a_{1}} = \ptest{12} \ket{a_{2}} = 0$.

The initial prepared state is 
$\ket{\Psi} = \ket{t} + \ket{a_{1}} + \ket{a_{2}} + \ket{a_{3}}$,
and the final post-selected state is 
$\bra{\Phi} = -\bra{t} + \bra{a_{1}} + \bra{a_{2}} + \bra{a_{3}}$.
Evaluating $\bra{\Phi} \ptest{ij} \ket{\Psi}$ for the various pairs, 
we find that {\em any} pair of particles would certainly be found to be
unrelated---i.e., in different boxes.  That is, $D = \{ (1,2), (2,3), (1,3) \}$.
This knowledge is inconsistent with any relation in $R$ and thus
constitutes quantum paradoxical knowledge.

\section{Penrose stairs}

Again we have three particles, each of which can occupy one of three
non-degenerate energy levels.  We are guaranteed that exactly
one particle is in each level.  There are thus six possible
states, each of which corresponds to a relation among the 
energies of the three particles.  For example, in state $\ket{123}$
we have $E_{1} < E_{2} < E_{3}$ while in state $\ket{312}$ we have
$E_{3} < E_{1} < E_{2}$.  The relations in $R$ are the total orderings
(in energy) on the particle set $\{ 1, 2, 3 \}$.
Each intermediate measurement $\ptest{ij}$ tests whether $E_{i} < E_{j}$.
Thus $\ptest{12}\ket{312} = \ket{312}$, but $\ptest{12} \ket{213} = 0$, etc.

As usual, the initial prepared state is 
\begin{equation} \label{eq-penrosepsi}
	\ket{\Psi} = \ket{123} + \ket{231} + \ket{312}  + \ket{132} + \ket{213} + \ket{321} .
\end{equation}
The final post-selected state is
\begin{equation}
	\bra{\Phi} = 2 \bra{123} + 2 \bra{231} + 2 \bra{312} - \bra{132} - \bra{213} - \bra{321} .
\end{equation}
That is, $\phi_{123} = \phi_{231} = \phi_{312} = 2$ and 
$\phi_{132} = \phi_{213} = \phi_{321} = -1$.  
The inner product $\amp{\Phi}{\Psi} = 3$.

Suppose the intermediate measurement is $\ptest{12}$.
\begin{equation}
	\bra{\Phi} \ptest{12} \ket{\Psi} = \phi_{123} + \phi_{321} + \phi_{132} = 3,
\end{equation}
which implies that this intermediate measurement would find
with certainty that $E_{1} < E_{2}$.  But since both $\ket{\Psi}$
and $\bra{\Phi}$ are invariant under cyclic permutation of the
particles, it immediately follows that a $\ptest{23}$ measurement
would yield that $E_{2} < E_{3}$ and that a $\ptest{31}$ 
measurement would yield $E_{3} < E_{1}$.  Including all possible
particle comparisons, our knowledge is
represented by $A = \{ (1,2), (2,3), (3,1) \}$ and 
$D = \{ (2,1), (3,2), (1,3) \}$.

This is plainly paradoxical.  The situation
is analogous to the famous Penrose stairs \cite{penrosestairs}, 
a visual paradox in which a staircase continually ascends in one
direction, but nevertheless appears to form a closed 
path.  
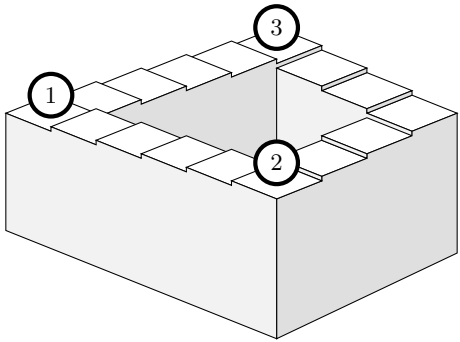
\begin{figure}
	\begin{center}
	\begin{tikzpicture}[scale=0.6]
		\myrhomb{0}{1.5}
		\myrhomb{-1}{1.2}
		\myrhomb{-2}{0.9}
		\myrhomb{-3}{0.6}
		\myrhomb{-4}{0.3}
		\myrhomb{-5}{0}
		\draw [fill=gray!25] (1,1.5)--(0,1.1)--(0,1.2)--(-1,0.8)--(-1,0.9)--(-2,0.5)--(-2,0.6)--(-3,0.2)--(-3,0.3)
		--(-4,-0.1)--(-4,-1.5)--(1,-1.5)--cycle;
		\myrhomb{1}{1.0}
		\myrhomb{2}{0.5}
		\myrhomb{3}{0}
		\draw [fill=gray!10] (0,1)--(1,0.6)--(1,0.5)--(2,0.1)--(2,0)--(3,-0.4)--(3,-2)--(0,-2)--cycle;
		\draw [fill=gray!25] (1,0.6)--(1,0.5)--(2,0.9)--(2,1)--cycle;
		\draw [fill=gray!25] (2,0.1)--(2,0)--(3,0.4)--(3,0.5)--cycle;
		\myrhomb{2}{-0.5}
		\myrhomb{1}{-1.0}
		\myrhomb{0}{-1.5}
		\draw  [fill=gray!25] (0,-1.9)--(1,-1.5)--(1,-1.4)--(2,-1)--(2,-0.9)--(3,-0.5)--(3,-0.4)--(4,0)--(4,-3.1)--(0,-5)--cycle;
		\draw [fill=gray!10] (1,-0.5)--(1,-0.6)--(2,-1)--(2,-0.9)--cycle;
		\draw [fill=gray!10] (0,-1)--(0,-1.1)--(1,-1.5)--(1,-1.4)--cycle;
		\myrhomb{-4}{-0.3}
		\myrhomb{-3}{-0.6}
		\myrhomb{-2}{-0.9}
		\myrhomb{-1}{-1.2}
		\myrhomb{0}{-1.5}
		\draw [fill=gray!10] (-6,0)--(-5,-0.4)--(-5,-0.3)--(-4,-0.7)--(-4,-0.6)--(-3,-1)--(-3,-0.9)
		--(-2,-1.3)--(-2,-1.2)--(-1,-1.6)--(-1,-1.5)--(0,-1.9)--(0,-5)--(-6,-2.6)--cycle;
		\node [circle,draw,fill=white,ultra thick] at (-5,0.4) {1};
		\node [circle,draw,fill=white,ultra thick] at (0,1.9) { 3};
		\node [circle,draw,fill=white,ultra thick] at (0,-1.1) {2};		
	\end{tikzpicture} 
	\end{center}
	\caption{Three particles on the Penrose stairs. \label{fig-penrose}}
\end{figure}
In Figure~\ref{fig-penrose}, we apparently
ascend the stairs (increasing potential energy) to go 
from 1 to 2, or from 2 to 3, or from 3 to 1.
This is precisely the type of quantum paradoxical knowledge 
that our preparation and post-selection provide.

\section{Paradoxical functions}

Our example of three particles distributed among three energy levels
can also represent a function relation.  Recall that relation $r$ is a 
function if, for every $i$, there exists a unique $j$ such that $(i,j) \in r$.
(We write this $r(i) = j$.)
The state we denoted by $\ket{ijk}$ represents a function on three
integers in which $r(1) = i$, $r(2) = j$ and $r(3) = k$.  We
restrict ourselves to one-to-one functions and prepare our system
in the state $\ket{\Psi}$ of Equation~\ref{eq-penrosepsi}.  At the 
intermediate time, we make one of two possible measurements, 
either determining whether $r(1) = 2$ or whether $r(1) = 3$.  Thus,
\begin{equation}
	\begin{split}
	\ptest{r(1)=2} \ket{\Psi} & = \ket{231} + \ket{213} , \\
	\ptest{r(1)=3} \ket{\Psi} & = \ket{312} + \ket{321} .
	\end{split}
\end{equation}
Finally, we post-select on the state
\begin{equation}
	\bra{\Phi} = - \bra{123} + \bra{231} + \bra{312} - \bra{132} + \bra{213} + \bra{321} .
\end{equation}
We observe that $\bra{\Phi} \ptest{r(1)=2} \ket{\Psi} = \bra{\Phi} \ptest{r(1)=3} \ket{\Psi} = \amp{\Phi}{\Psi} = 2$.
Preparation and post-selection have thus yielded the paradoxical knowledge
that $r(1) = 2$ and $r(1) = 3$, contradicting the uniqueness of the 
function value $r(1)$.

\section{Unconditionally paradoxical knowledge}

So far, our examples of quantum paradoxical knowledge
require post-selection using a particular 
outcome for the final measurement.  Can we construct
an example in which {\em every} outcome of the final
measurement yields paradoxical knowledge?

We can.  In fact, the original quantum pigeonhole
example works in this way, as pointed out in \cite{qpigeonhole}.  
If the final measurement resolves the $\{ \ket{+i},\ket{-i} \}$ 
basis for each particle, there are 8 possible outcomes.  
Two of these yield violations of the pigeonhole principle, 
in which no two particles can be found together.  
The remaining six outcomes lead to a different type 
of paradoxical knowledge.  For example, the result
$\ket{+i_1,+i_2,-i_3}$ yields $A = \{ (1,3),(2,3) \}$ and
$D = \{ (1,2) \}$.  Our knowledge of the togetherness
relation is not transitive, and thus is paradoxical.

\begin{figure}
	\begin{center}
	\begin{tabular}{cc}
        	\begin{tikzpicture}[scale=1.0]
        		\node (A) at (0,0) {$1$};
        		\node (B) at (0,1.0) {$2$};
        		\node (C) at (-1,-0.5) {$3$};
		\node (D) at (1,-0.5) {$4$};
        		\node (R) at (0,-1) {$r_1$};
        		\draw (A) -- (B);
        		\draw (A) -- (C);
        		\draw (A) -- (D);
		\draw (-1.25,-1.25) rectangle (1.25,1.25);
	\end{tikzpicture} 
	& 
		\begin{tikzpicture}[scale=1.0]
        		\node (A) at (0,0) {$2$};
        		\node (B) at (0,1.0) {$3$};
        		\node (C) at (-1,-0.5) {$4$};
		\node (D) at (1,-0.5) {$1$};
        		\node (R) at (0,-1) {$r_2$};
        		\draw (A) -- (B);
        		\draw (A) -- (C);
        		\draw (A) -- (D);
		\draw (-1.25,-1.25) rectangle (1.25,1.25);
	\end{tikzpicture} 
	\\
		\begin{tikzpicture}[scale=1.0]
        		\node (A) at (0,0) {$3$};
        		\node (B) at (0,1.0) {$4$};
        		\node (C) at (-1,-0.5) {$1$};
		\node (D) at (1,-0.5) {$2$};
        		\node (R) at (0,-1) {$r_3$};
        		\draw (A) -- (B);
        		\draw (A) -- (C);
        		\draw (A) -- (D);
		\draw (-1.25,-1.25) rectangle (1.25,1.25);
	\end{tikzpicture} 
	& 
		\begin{tikzpicture}[scale=1.0]
        		\node (A) at (0,0) {$4$};
        		\node (B) at (0,1.0) {$1$};
        		\node (C) at (-1,-0.5) {$2$};
		\node (D) at (1,-0.5) {$3$};
        		\node (R) at (0,-1) {$r_4$};
        		\draw (A) -- (B);
        		\draw (A) -- (C);
        		\draw (A) -- (D);
		\draw (-1.25,-1.25) rectangle (1.25,1.25);
	\end{tikzpicture} 
	\end{tabular}
	\end{center}
	\caption{Star graphs on 4 particles.  \label{fig-4star}}
\end{figure}
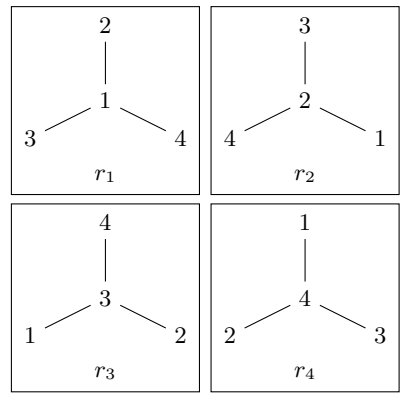

Let us construct our own example of this sort of 
unconditionally paradoxical knowledge.  The relations
in $R$ are represented by star graphs on the $N$
particles.  A single particle lies in the center, and it
is related to each of the others, but none of the others
are related to one another.  We designate a particular star
graph by identifying the center particle.
There are $N$ such relations in $R$.  The ones for
$N=4$ are shown in Figure~\ref{fig-4star}.
Each relation $r_{i} \in R$ corresponds to a state $\ket{r_{i}}$
of the system.  The initial state is again a uniform superposition
of these, and the final measurement employs the following
orthogonal basis states in the 4-dimensional Hilbert space (given in
bra form):
\begin{equation}
	\begin{split}
		\bra{\Phi_{1}} & = - \bra{r_{1}} + \bra{r_{2}} + \bra{r_{3}} + \bra{r_{4}} \\
		\bra{\Phi_{2}} & = + \bra{r_{1}} - \bra{r_{2}} + \bra{r_{3}} + \bra{r_{4}} \\
		\bra{\Phi_{3}} & = + \bra{r_{1}} + \bra{r_{2}} - \bra{r_{3}} + \bra{r_{4}} \\
		\bra{\Phi_{4}} & = + \bra{r_{1}} + \bra{r_{2}} + \bra{r_{3}} - \bra{r_{4}} 
	\end{split} .
\end{equation}
Every possible outcome provides paradoxical knowledge.  For example,
if we obtain result $\ket{\Phi_{1}}$, we find that particle 1 is isolated
and the remaining three particles are connected, yielding
$A = \{ (2,3), (2,4), (3,4) \}$ and $D = \{ (1,2), (1,3), (1,4) \}$.
For each final result, we have one isolated particle and
three mutually connected particles.
No such graph is in $R$, and so every one constitutes quantum
paradoxical knowledge.

\section{Star graphs and map coloring}

The star graph example in the previous section has a conditional 
extension to more particles.  Suppose we have $N>3$ particles that are
related in one of the $N$ possible star graphs, each labeled as before
by its central particle.  The initial state $\ket{\Psi}$ is a
uniform superposition of the $N$ terms for these relations in $R$.  
We post-select by a final state $\bra{\Phi}$ such that
\begin{equation}
	\begin{split}
	\phi_{1} & = \ldots \phi_{N-1} = 1 ,\\
	\phi_{N} & = - (N-3) ,
	\end{split}
\end{equation}
so that $\amp{\Phi}{\Psi} = 2$.  If both $i$ and $j$ are in the range
$1, \ldots, N-1$, we find that $\bra{\Phi} \ptest{ij} \ket{\Psi} = 2$ also,
so that $i$ and $j$ are definitely related.  That is, our post-selected
knowledge of the relation entails that each pair from the first $N-1$ particles
is certainly related.  Our knowledge includes a {\em clique} of size $N-1$,
which is paradoxical given $R$.

This type of paradoxical knowledge has a notable implication.  Every
star graph is planar, so that it can represent adjacency relations 
among regions in a 2-D planar map.  
We are, in effect, distributing our particles among these regions.  
A celebrated theorem \cite{fourcolor} guarantees that, 
given no more than four distinct colors, we can always
assign colors to the regions of a planar map so that adjacent regions 
are colored differently.  However, a clique of $N-1$ particles
cannot be so colored with fewer than $N-1$ colors.
If $N \geq 6$, therefore, our quantum paradoxical knowledge violates the 
four-color theorem.  In quantum map coloring, four colors do {\em not} suffice.

It is worth noting that Tang \cite{tang2022} has used a very different framework
to connect graph coloring to quantum pigeonhole-like paradoxes involving
$n$ entangled qubits.  These paradoxes amount to Hardy-like constructions
for excluding local hidden variable descriptions of entangled states.

\section{Beyond binary relations}

A natural generalization of our formalism would be to extend it
to ternary (or $n$-ary) relations among the $N$ particles.  For
example, suppose we have three particles designated $a$, $b$
and $c$, each of which may have energies $0$, $1$ or $2$ in
appropriate units.  We require that $E_{a} + E_{b} = E_{c}$.
This is a ternary relation among the particle states.  The
relevant states may be written
\begin{equation}
	\ket{000}, \ket{101}, \ket{011}, \ket{112}, \ket{202}, \mbox{ and }\ket{022} .
\end{equation}
Our initial state $\ket{\Psi}$ is a uniform superposition of these.
At the intermediate time, we measure whether $E_{a} = 1$,
whether $E_{b} = 1$ or whether $E_{c} = 2$.  We post-select
on the state
\begin{equation}
	\bra{\Phi} = 2 \bra{101} + 2 \bra{011} + 2 \bra{112} - \bra{202} - \bra{202} .
\end{equation}
It is easy to see that $\bra{\Phi} \ptest{E_{a}=1} \ket{\Psi}
= \bra{\Phi} \ptest{E_{b}=1} \ket{\Psi}
= \amp{\Phi}{\Psi} = 4$, but
$\bra{\Phi} \ptest{E_{c}=2} \ket{\Psi} = 0$.
We thus have the quantum paradoxical knowledge that
$E_{a}=1$ and $E_{b} = 1$, but $E_{c} = E_{a} + E_{b} \neq 2$.

\section{Remarks}

We have generalized the quantum pigeonhole paradox to a larger 
class of paradoxes involving many different types of relations
between particles.  These paradoxes can be quite simple.  For example,
suppose $R$ includes star graphs on just three particles.  In every such
graph, no particle is isolated from the others.  With a post-selected
state $\bra{\Phi} = - \bra{r_{1}} + \bra{r_{2}} + \bra{r_{3}}$, however,
we find that $\bra{\Phi} \ptest{12} \ket{\Psi} = \bra{\Phi} \ptest{13} \ket{\Psi} = 0$.
Particle 1 can never be found linked to either of the other two; it is
paradoxically isolated.

The quantum pigeonhole paradox is particularly appealing, because it involves
straightforward position observables.  Both the preparation and 
post-selection are actually product states of the particles.  This
has motivated experimental implementations of the pigeonhole scenario 
\cite{nmr-pigeonhole,neutron-pigeonhole,photon-pigeonhole}.
Our general framework, being more abstract, is somewhat further removed from
a simple experimental set-up.  The Penrose stairs paradox, for 
instance, employs entangled preparation and post-selection states
for three particles in three energy levels.  However, it should be possible to realize 
all of our examples as quantum computations involving a modest number
of qubits.

Not every set $R$ of relevant binary relations on $N$ particles can lead to
quantum paradoxical knowledge.  For instance, it is not hard to show
that we must have $\numberin{R} \geq 3$ for our framework to yield
a paradox.  The three-particle example above is minimal in this sense.
At other end of the spectrum, if the set $R$ includes every relation, 
then any pair $[ A,D ]$ of disjoint affirmable and deniable sets 
will be consistent 
with some element $r \in R$, and thus cannot be paradoxical.
Therefore, $R$ must be neither too small nor too large.  
For what sets of relations $R$ on $N$ particles can
generalized quantum paradoxical knowledge arise?


\begin{thebibliography}{9}
%
\bibitem{qpigeonhole}  Aharonov, Y., Colombo, F., Popescu, S., and Tollaksen, J., 
	Quantum violation of the pigeonhole principle and the nature of quantum
	correlations.  {\em PNAS} {\bf 113}, 532--535 (2016).
%
\bibitem{penrosestairs}  Penrose, L. S. and Penrose, R., Impossible objects: A special
	type of visual illusion.  {\em British Journal of Psychology} {\bf 49}, 31--33 (1958).
%
\bibitem{fourcolor} Appel, K. and Haken, W., Every Planar Map Is Four Colorable.
	{\em Bull. Amer. Math. Soc.} {\bf 82}, 711--712 (1976).
%
\bibitem{tang2022}  Tang, W., Hardy-like quantum pigeonhole and the projected-coloring
	graph state.  {\em Phys. Rev. A} {\bf 105}, 032457 (2022).
%
\bibitem{nmr-pigeonhole} Anjusha, V. S., Hegde, S. S., Mahesh, T. S., NMR 
	investigation of the quantum pigeonhole effect.  {\em Phys. Lett. A}
	{\bf 380}, 577--580 (2016).
%
\bibitem{neutron-pigeonhole}  Waegell, M. et al., Confined contextuality in neutron
	interferometry:  Observing the quantum pigeonhole effect.
	{\em Phys. Rev. A} {\bf 96}, 052131 (2017).
%
\bibitem{photon-pigeonhole} Chen, M.-C. et al., Experimental demonstration of 
	quantum pigeonhole paradox.  {\em PNAS} {\bf 116}, 1549--1552 (2019).
%
\end{thebibliography}
\end{document}